\documentclass[twocolumn,aps,prl]{revtex4}

\usepackage{graphicx}
\usepackage{amsmath}

\newcommand{\eq}[1]{(\ref{#1})}
\newcommand{\ket}[1]{\left| #1 \right\rangle}

\newcommand{\spp}[1]{\sigma_{+ #1}}
\newcommand{\smm}[1]{\sigma_{- #1}}

\newcommand{\ad}{a^\dag}
\newcommand{\aop}{a}

\def\be{\begin{equation}}
\def\ee{\end{equation}}

\begin{document}

\title{Sideband Transitions and Two-Tone Spectroscopy\\
of a Superconducting Qubit Strongly Coupled to an On-Chip Cavity}

\author{A.~Wallraff}
\altaffiliation{Department of Physics, ETH Zurich, CH-8093
Z{\"u}rich, Switzerland} \affiliation{Departments of Applied Physics
and Physics, Yale University, New Haven, CT 06520}
\author{D.~I.~Schuster}
\affiliation{Departments of Applied Physics and Physics, Yale
University, New Haven, CT 06520}
\author{A.~Blais}
\altaffiliation{D\'{e}partement de Physique et R\'{e}groupement
Qu\'{e}b\'{e}cois sur les Mat\'{e}riaux de Pointe, Universit\'{e} de
Sherbrooke, Sherbrooke, Qu\'{e}bec, Canada, J1K 2R1}
\affiliation{Departments of Applied Physics and Physics, Yale
University, New Haven, CT 06520}
\author{J.~M.~Gambetta}
\affiliation{Departments of Applied Physics and Physics, Yale
University, New Haven, CT 06520}
\author{J.~Schreier}
\affiliation{Departments of Applied Physics and Physics, Yale
University, New Haven, CT 06520}
\author{L.~Frunzio}
\affiliation{Departments of Applied Physics and Physics, Yale
University, New Haven, CT 06520}
\author{M.~H.~Devoret}
\affiliation{Departments of Applied Physics and Physics, Yale
University, New Haven, CT 06520}
\author{S.~M.~Girvin}
\affiliation{Departments of Applied Physics and Physics, Yale
University, New Haven, CT 06520}
\author{R.~J.~Schoelkopf}
\affiliation{Departments of Applied Physics and Physics, Yale
University, New Haven, CT 06520}
\date{\today}

\begin{abstract}
Sideband transitions are spectroscopically probed in a system
consisting of a Cooper pair box strongly but non-resonantly coupled
to a superconducting transmission line resonator. When the Cooper
pair box is operated at the optimal charge bias point the symmetry
of the hamiltonian requires a two photon process to access
sidebands. The observed large dispersive ac-Stark shifts in the
sideband transitions induced by the strong non-resonant drives agree
well with our theoretical predictions. Sideband transitions are
important in realizing qubit-photon and qubit-qubit entanglement in
the circuit quantum electrodynamics architecture for quantum
information processing.
\end{abstract}

\maketitle

A promising route towards the implementation of a scalable solid
state quantum information processor \cite{Nielsen00} is based on
superconducting quantum electronic circuits. The basic concepts of
creating coherent quantum two-level systems (qubits) from
superconducting circuit elements such as inductors, capacitors and
Josephson junctions (an ideal, non-linear inductor) are well
understood \cite{Devoret04} and a variety of qubits have been
implemented in a wide range of architectures \cite{Wendin05}. The
realization of controlled qubit interactions
\cite{Berkley03,Pashkin03,Majer04,Chiorescu04,McDermott04} and the
implementation and characterization of two-qubit gate operations
\cite{Yamamoto03,Steffen06} is a main objective of current research.
Many implementations are, however, limited to static nearest
neighbor couplings. By coupling individual qubits to a common
harmonic oscillator mode used as a bus \cite{Makhlin99,Blais04,Blais06}
non-local interactions can be mediated between very distant qubits,
a much more versatile and scalable approach for quantum information
processing.

A promising realization of such a system is based on a set of Cooper
pair boxes strongly coupled to a high quality transmission line
resonator \cite{Blais04}. In this circuit quantum electrodynamics
(QED) architecture \cite{Blais04}, it has been demonstrated
spectroscopically that a single photon can be exchanged coherently
between a superconducting cavity and an individual qubit
\cite{wallraff04c} in a \emph{resonant} process known as the vacuum
Rabi mode splitting. This resonant process has also been observed as
time-resolved oscillations in a persistent current qubit coupled to
a lumped element oscillator \cite{Chiorescu04,Johansson06}. This
feat is essential for non-local qubit coupling and for quantum
communication as it allows to transfer quantum information from
stationary qubits to photons used as `flying' qubits that may enable
hybrid quantum information systems \cite{Andre06}.

\begin{figure}[!bp]
\includegraphics[width = 1.0 \columnwidth]{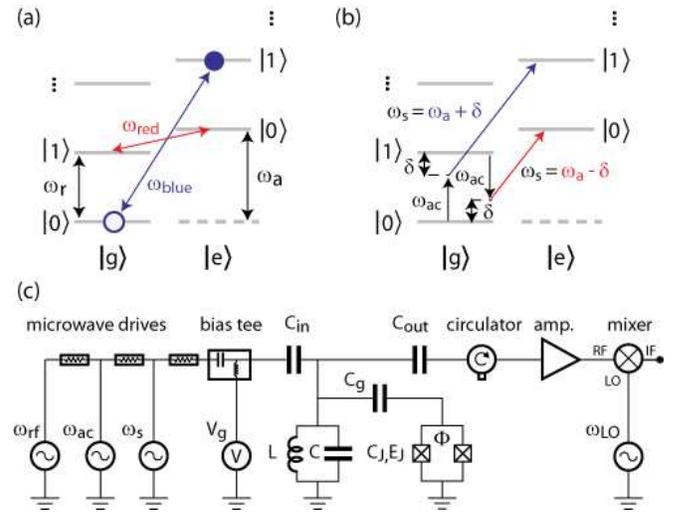}
\caption{(color online) (a) Dispersive dressed states energy level
diagram for qubit states $|g\rangle$ and $|e\rangle$ with separation
$\omega_a$ and single mode cavity photon states $|n = 0, 1, 2
...\rangle$ with separation $\omega_r$. Red ($\omega_{\rm{red}}$)
and blue ($\omega_{\rm{blue}}$) side band transitions are indicated.
(b) Two-tone scheme for sideband transitions. Fixed ac-tone
$\omega_{\rm{ac}}$ detuned by $\delta$ from cavity combined with
spectroscopy tone $\omega_{\rm{s}} = \omega_{\rm{a}} \pm \delta$
induces sideband transitions. (c) Measurement setup for two-tone
spectroscopy, compare with setups in
Refs.~\cite{wallraff04c,wallraff05}. An extra phase coherent
microwave source at the ac-tone is applied to the cavity input.}
\label{fig:2ToneSideBand}
\end{figure}

In the circuit QED architecture high fidelity qubit control has been
demonstrated and the strong \emph{non-resonant} (or dispersive)
coupling of individual qubits to cavity photons has been
successfully employed for high visibility readout of the qubit state
\cite{Schuster05,wallraff05}. In the dispersive regime, when the
transition frequency $\omega_a$ between the qubit ground state
$|g\rangle$ and excited state $|e\rangle$ is detuned by an amount
$\Delta = \omega_a - \omega_r$ from the cavity frequency $\omega_r$,
the resonant qubit-photon interaction is suppressed. Nevertheless,
\emph{sideband transitions}, see Fig.~\ref{fig:2ToneSideBand}a, may
be used to transfer a qubit state to a photon state, also see
Ref.~\cite{liu05a}. Sideband transitions are induced by driving the
coupled qubit-cavity system at its sum
$\omega_{\rm{blue}}=\omega_{\rm{a}} + \omega_{\rm{r}}$ (\emph{blue
sideband}) or difference frequencies
$\omega_{\rm{red}}=\omega_{\rm{a}} - \omega_{\rm{r}}$ (\emph{red
sideband}). Similarly sidebands have been accessed in a system in
which a qubit is coupled to a SQUID oscillator also used for readout
\cite{Chiorescu04}. Using such sideband transitions, entanglement
between a qubit and the resonator could potentially be generated, a
process essential for realizing non-local gate operations in a set
of qubits coupled to a cavity.

In this letter, we demonstrate an approach to access the red and
blue sideband transitions of a Cooper pair box biased at its optimal
point and coupled to a cavity using two microwave tones at different
frequencies. We spectroscopically probe the sideband transitions and
explain the observed transition frequencies induced by two
independent non-resonant drives.

\begin{figure}[!b]
\includegraphics[width = 1.0 \columnwidth]{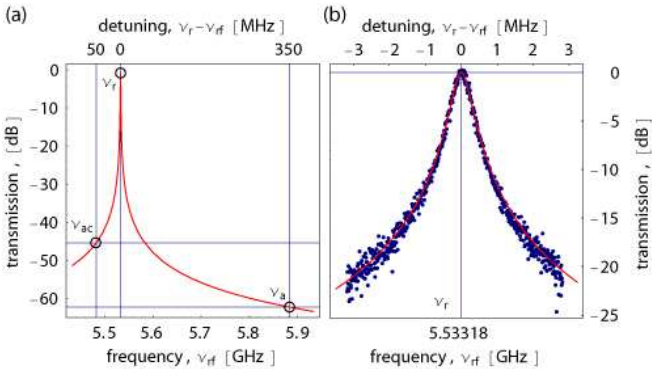}
\caption{(color online) (a) Calculated lorentzian cavity
transmission spectrum over large frequency range. The cavity drives
at frequencies $\nu_{\rm{s}} \approx \nu_{\rm{a}}$, $\nu_{\rm{rf}}
\approx \nu_{\rm{r}}$ and $\nu_{\rm{ac}}$ and the cavity
transmission at these frequencies are indicated. (b) Measured cavity
spectrum (dots) and fit (line) around cavity frequency.}
\label{fig:resonatorlineCombined}
\end{figure}

We consider a Cooper pair box \cite{Bouchiat98} with Josephson
energy $E_{\rm{J,max}} = 12 \, \rm{GHz}$ and single electron
charging energy $E_{\rm{c}} = 4.75 \, \rm{GHz}$ biased at its
optimal point \cite{Vion02} as an ideal two-level system with a
ground state $|g\rangle$, an excited state $|e\rangle$ and bare
transition frequency $\omega_{\rm{a}} = 5.8 \,\rm{GHz}$. The Cooper
pair box is coupled with strength $g/2\pi = 17 \, \rm{MHz}$ to a
harmonic oscillator with states $|0\rangle, |1\rangle, ...
|n\rangle$ realized as a single mode of a transmission line
resonator \cite{Frunzio05} with frequency $\omega_r/2\pi = 5.5 \,
\rm{GHz}$ and decay rate $\kappa/2\pi = 0.5 \, \rm{MHz}$. The strong
coupling limit of cavity QED is realized in this system
\cite{wallraff04c} and its dynamics are described by the
Jaynes-Cummings hamiltonian \cite{Blais04}. Using applied magnetic
flux, the split Cooper pair box is detuned from the cavity by
$\Delta = \omega_{\rm{a}}-\omega_{\rm{r}} \approx 2\pi \, 350 \,
\rm{MHz} \gg g$.

The qubit transition frequency $\omega_{\rm{a}}/2\pi = 5.877 \,
\rm{GHz}$ is found by applying a spectroscopy microwave tone at
frequency $\omega_{\rm{s}}$ to the cavity and measuring the phase
shift of a microwave beam applied to the system at the cavity
frequency $\omega_{\rm{r}}$, as demonstrated in
Ref.~\onlinecite{Schuster05}. Using this single frequency scheme
however, the red and blue sidebands cannot be observed because the
single photon transitions are to first order forbidden when the
Cooper pair box is biased at charge degeneracy. This can be seen by
introducing the parity operator $P= e^{-i\pi a^\dag a}\sigma_z$ and
recognizing that states involved in the sideband transitions are of
equal parity while the term in the Hamiltonian responsible to drive
these transitions is of odd parity~\cite{Blais06,liu05a,liu05b}.
Away from charge degeneracy, the sideband transitions are allowed,
but at the expense of reduced coherence times due to the larger
sensitivity of the qubit to charge noise \cite{Ithier05}.

Going to second order in the drive Hamiltonian, the above
considerations mean that two-photon sideband transitions are allowed
at charge degeneracy.  One way to drive these two-photon transitions
is simply to choose the drive frequency to be $\omega_{\rm{red,
blue}}/2$. The red sideband, however, is strongly detuned from the
resonator and it is difficult to drive the transition at the
required rate because of the filtering due to the cavity at large
detunings, also see Fig.~\ref{fig:resonatorlineCombined}.

\begin{figure}[!b]
\includegraphics[width = 1.0 \columnwidth]{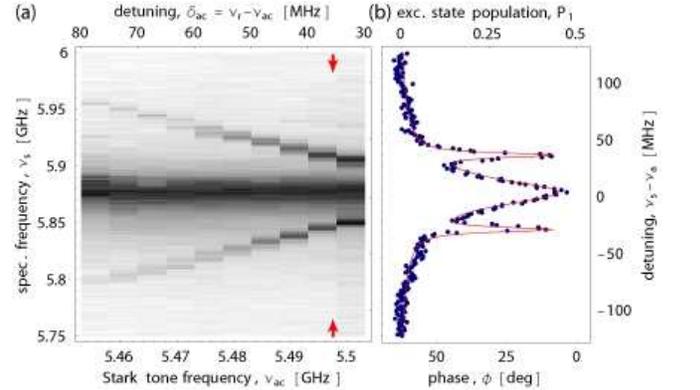}
\caption{(color online) (a) Density plot of measured cavity phase
shift $\phi$ (white : $70 \,\rm{deg}$, black : $0 \, \rm{deg}$)
\textsl{vs.} $\omega_{\rm{s}}$ and $\omega_{\rm{ac}}$.
(b) Measured $\phi$ \textsl{vs.} $\omega_{\rm{s}}$ for $\delta/2\pi
= 50 \, \rm{MHz}$ (dots) as indicated by arrows in (a) and fit to
linear combination of three independent Lorentzian line shapes
(solid line).} \label{fig:232slice}
\end{figure}

As a result, here, we choose to drive sideband transitions at charge
degeneracy with two photons of frequencies $\omega_{\rm{ac}}$ and
$\omega_{\rm{s}}$ that can be selected freely with the constraint
that their sum or difference must match the desired sideband
$\omega_{\rm{red,blue}}=\omega_{\rm{ac}} \pm \omega_{\rm{s}}$. In
particular, an off-resonant ac-Stark drive at fixed frequency
$\omega_{\rm{ac}} = \omega_{\rm{r}} - \delta$ is chosen at a small
detuning $\delta$ from the cavity frequency, see
Fig.~\ref{fig:2ToneSideBand}b. The power coupled into the cavity for
a fixed external drives scales as $1/\delta^2$ and thus is largest
for small values of $\delta$. A second microwave drive applied at
frequency $\omega_{\rm{s}} \approx \omega_{\rm{a}} \pm \delta$ then
induces two-photon blue or red sideband transitions, respectively,
as shown in Fig.~\ref{fig:2ToneSideBand}b.  The effective
Hamiltonians describing these processes are given by~\cite{Blais06}
\be
\begin{split}
H_\mathrm{2R} \approx \frac{g}{4}
\left(\frac{\Omega_{\rm{ac}}}{\omega_{\rm{a}}-\omega_{\rm{ac}}}\right)
\left(\frac{\Omega_{\rm{s}}}{\omega_{\rm{a}}-\omega_{\rm{s}}}\right)
\left( \ad\smm{}+\aop\spp{} \right),
\\
H_\mathrm{2B} \approx \frac{g}{4}
\left(\frac{\Omega_{\rm{ac}}}{\omega_{\rm{a}}-\omega_{\rm{ac}}}\right)
\left(\frac{\Omega_{\rm{s}}}{\omega_{\rm{a}}-\omega_{\rm{s}}}\right)
\left( \ad\spp{}+\aop\smm{} \right),
\end{split}
\ee
where $  \Omega_{\rm{ac,s}}$ are Rabi frequencies given by
\be
\Omega_{\rm{ac,s}} = \frac{2 g \epsilon_{\rm{ac,s}}}{\omega_r -
\omega_{\rm{ac,s}}}, \label{eq:specdrive} \ee with the drive
amplitudes $\epsilon_{\rm{ac,s}}$ expressed as frequencies, also see
Eq.~(\ref{eq:photonnumber}).

Figure \ref{fig:232slice}b shows the cavity phase shift $\phi$ in
the presence of a fixed power and fixed frequency ac-Stark tone
chosen at a detuning of $\delta/2\pi =  35 \, \rm{MHz}$ in response
to a spectroscopic drive scanned around the qubit transition
frequency. The main qubit transition at $\omega_{\rm{a}}$ and the
two sidebands at $\omega_{\rm{a}} \pm \delta$ are clearly observed.
We also note that the sideband transitions can be saturated at
suitably chosen drive powers indicating that sideband oscillations
should be observable in future experiments. In
Fig.~\ref{fig:232slice}a, the side band transitions are shown
varying the detuning $\delta$ but keeping the drive amplitudes
fixed. In this parameter range, the sideband frequenies scale as
expected to first order linearly with the detuning $\delta$ and the
amplitudes of the peaks decrease with detuning as the effective
photon number in the cavity decreases with $\delta$ due to the
cavity filtering.

\begin{figure}[!bp]
\includegraphics[width = 0.90 \columnwidth]{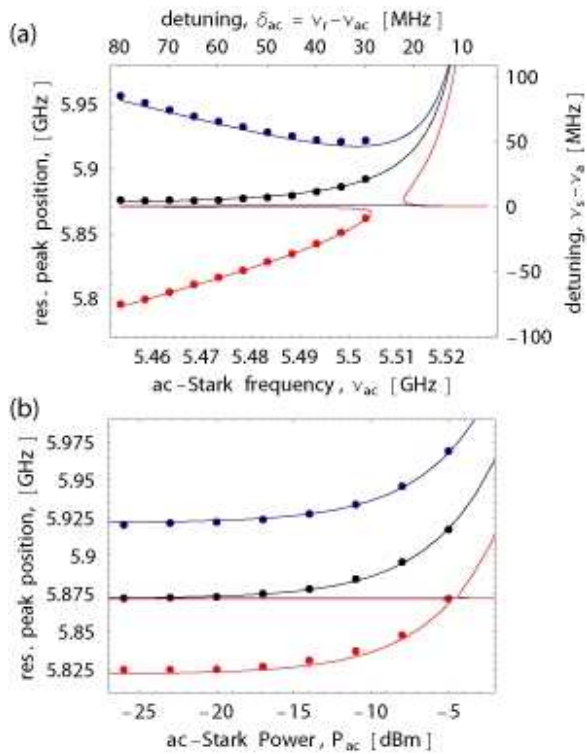}
\caption{(color online) (a) Two-tone red and blue sidebands
(red/blue dots) and fundamental (black dots) qubit transition
frequencies for fixed $P_{\rm{s}}$ and $P_{\rm{ac}}$ varying
$\omega_{\rm{ac}}$. Lines are fits to
Eqs.~(\ref{eq:renormqubit}-\ref{eq:rescond}) (b) For fixed
$\omega_{\rm{ac}}$,  and varying $P_{\rm{ac}}$. Measurements are
done at fixed measurement frequency $\omega_{\rm{rf}}=
\omega_{\rm{r}}$ and power $P_{\rm{rf}}$ weakly populating the
resonator with $n_{\rm{rf}} \approx 2$ photons.}
\label{fig:238a254a258}
\end{figure}

At higher drive amplitudes, however, large shifts of the sideband
frequencies from $\omega_a\pm\delta$ become apparent. In
Fig.~\ref{fig:238a254a258}a, the center and sideband transition
frequencies are indicated as extracted from the peak positions of
Lorentzian line fits, similar to those in Fig.~\ref{fig:232slice}b,
for an ac-Stark tone power increased by a factor of $4$ in
comparison to the measurement presented in Fig.~\ref{fig:232slice}.
This data shows that all transitions are shifted to higher
frequencies as the detuning $\delta$ decreases, an effect that can
be explained considering the ac-Stark shifts \cite{Schuster05}
induced in the qubit transition frequency by the off resonant
drives. As both drives at $\omega_{\rm{ac}}$ and $\omega_{\rm{s}}$
are detuned from the qubit transition by $\delta$ and $\Delta +
\delta$ they induce dispersive shifts  in the qubit frequency
\begin{equation}
    \tilde{\omega}_{\rm{a}} =
    \omega_a
    + \frac{1}{2} \frac{\Omega^2_{\rm{ac}}}{\omega_{\rm{a}}-\omega_{\rm{ac}}}
    + \frac{1}{2} \frac{\Omega^2_{\rm{s}}}{\omega_{\rm{a}}-\omega_{\rm{s}}} \, ,
    \label{eq:renormqubit}
\end{equation}
that depend on the drive strengths $\epsilon_{\rm{ac,s}}$ and
frequencies $\omega_{\rm{ac,s}}$.  The shift due to the additional
drive used to measure the qubit population is taken into account in
the same way~\cite{Schuster05}. These off-resonance ac-Stark shifts
are important at large drive amplitudes, i.e.~when the qubit
transition is saturated, and small detunings. It is important to
note that any frequency multiplexed scheme used to address multiple
qubits is affected by such shifts. However, the effect can be
compensated for by appropriately readjusting the qubit drive
frequencies.

At a fixed ac-Stark tone, the blue and red sideband frequencies are
then determined by the nonlinear equation
\begin{equation}
    \omega_{\rm{s}} = \tilde{\omega}_{\rm{a}} \pm \delta
    \label{eq:rescond}
\end{equation}
for $\omega_s$ that allows one to accurately fit the center
frequencies of both the fundamental and the sidebands with the same
set of parameters, see Fig.~\ref{fig:238a254a258}.

The drive amplitudes required for the fits using
Eq.~(\ref{eq:specdrive}) were calibrated using the ac-Stark shift
of the qubit transition induced by the measurement beam resonant
with the cavity frequency $\omega_{\rm{r}}$. At this frequency an
input power of  $P^{(1)}_{\rm{rf}} = -28 \, \rm{dBm}$ corresponds
to an average cavity photon occupation number of $n = 1$. The
photon number $n$ is related to the drive amplitude $\epsilon$
\begin{equation}\label{eq:photonnumber}
    n = \frac{\epsilon^2}{\Delta^2 + (\kappa/2)^2} \, ,
\end{equation}
depending on the detuning from the cavity $\Delta$ and the cavity
decay rate $\kappa$. Having accurately measured $\kappa$ (see
Fig.~\ref{fig:resonatorlineCombined}b) and knowing $\Delta$, the
values for $\epsilon$ are accurately and consistently determined for
all fits.

\begin{figure}[tbp]
\includegraphics[width = 0.80 \columnwidth]{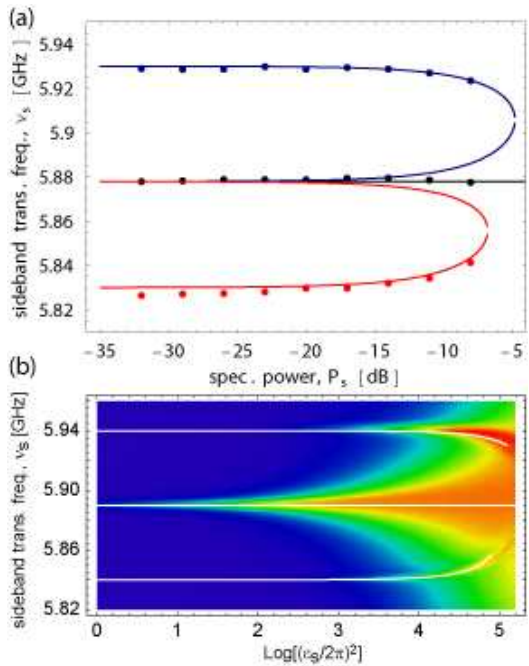}
\caption{(color online) (a) As Fig.~\ref{fig:238a254a258}b but for
fixed $\omega_{\rm{ac}}$, $P_{\rm{ac}}$ and varying $P_{\rm{s}}$.
(b) Master equation simulation of the qubit inversion $\langle
\sigma_z\rangle$ as a function of spectroscopy frequency
$\omega_\mathrm{s}/2\pi$ and the spectroscopy power
$\log(\epsilon_\mathrm{s}/2\pi)^2$ on a logarithmic scale. Blue:
$\langle \sigma_z\rangle = -1$;  Red: $\langle \sigma_z\rangle = 0$.
The white lines are obtained from Eqs.~\eq{eq:renormqubit} and
\eq{eq:rescond}.  The parameter values used in the simulation are
those quoted in the text. } \label{fig:calculations}
\end{figure}

For the sideband measurements presented in
Fig.~\ref{fig:238a254a258}b, we have chosen a fixed detuning of
$\delta/2\pi = 50 \, \rm{MHz}$ and have varied the ac-Stark tone
power over two orders of magnitude and kept the dispersive shifts
due to the spectroscopy drive minimal by keeping the drive amplitude
low. As observed, the dispersive shifts in the qubit frequency with
drive power can be much larger than the qubit line widths and are
well described by Eqs.~(\ref{eq:renormqubit}-\ref{eq:rescond}), see
lines in Fig.~\ref{fig:238a254a258}. As the frequencies of the two
tones are smaller than both the cavity and the qubit frequency all
lines are shifted to higher frequencies, see
Eq.~(\ref{eq:renormqubit}).

In the case that the spectroscopy drive amplitude dominates the
dispersive shifts, the red sideband is shifted to higher
frequencies, whereas the blue sideband is shifted to lower
frequencies, since the detuning of the spectroscopy drive from the
bare qubit transition frequency has different signs. This situation
is analyzed in Fig.~\ref{fig:calculations}, where the spectroscopy
power is varied by 3 orders of magnitude at fixed detuning of
$\delta = 50 \, \rm{MHz}$. Again the sideband transition frequencies
are predicted well by our analysis. As shown in
Fig.~\ref{fig:calculations}b, these results also agree well with
numerical simulation of the system's master equation in the
Born-Markov approximation~\cite{Blais06}.

We have demonstrated that sideband transitions can be driven in a
Cooper pair box coupled to a transmission line resonator and biased
at the optimal point using two photons of different frequencies.
Large shifts in the qubit level separation are observed and can be
explained and predicted considering the ac-Stark shifts induced by
the non-resonant drives. These are particularly important at high
drive amplitudes that are required for short pulses that can be
potentially used to entangle a qubit with a cavity photon and to
generate qubit-qubit entanglement. A protocol for entangling qubits
could follow these lines. We would start out with both the cavity
and a detuned qubit in their ground state $|0,g\rangle$ and apply a
$\pi$ pulse to the qubit to generate the state $|0,e\rangle$. With a
$\pi/2$ red sideband pulse on the coupled system we would generate a
maximally entangled cavity-qubit state of the form $|0,e\rangle +
|1,g\rangle$. This entaglement would then be transferred to generate
a qubit-qubit entangled state. Applying a $\pi$ pulse on the red
sideband of a second qubit would prepare a final state of the form
$\ket{0}\otimes(|e,g\rangle + |g,e\rangle)$, which is a Bell state
of two superconducting qubits and leaves the cavity in its ground
state and completely unentangled from the qubits. This scheme would
be an off-resonant implementation of the generation of entangled
qubit pairs inspired by the protocol demonstrated for Rydberg atoms
\cite{raimond01} and trapped ions~\cite{schmidt-kaler03}. We also
note that single microwave photon sources could be realized and
cavity Fock states could be prepared in circuit QED using this
sideband scheme.

\begin{acknowledgments}
This work was supported in part by the NSA under ARO contract
W911NF-05-1-0365, and the NSF under ITR 0325580 and DMR-0603369, and
the W. M. Keck Foundation. AB was partially supported by NSERC,
FQRNT and CIAR.
\end{acknowledgments}


\begin{thebibliography}{24}
\expandafter\ifx\csname
natexlab\endcsname\relax\def\natexlab#1{#1}\fi
\expandafter\ifx\csname bibnamefont\endcsname\relax
  \def\bibnamefont#1{#1}\fi
\expandafter\ifx\csname bibfnamefont\endcsname\relax
  \def\bibfnamefont#1{#1}\fi
\expandafter\ifx\csname citenamefont\endcsname\relax
  \def\citenamefont#1{#1}\fi
\expandafter\ifx\csname url\endcsname\relax
  \def\url#1{\texttt{#1}}\fi
\expandafter\ifx\csname urlprefix\endcsname\relax\def\urlprefix{URL
}\fi \providecommand{\bibinfo}[2]{#2}
\providecommand{\eprint}[2][]{\url{#2}}

\bibitem[{\citenamefont{Nielsen and Chuang}(2000)}]{Nielsen00}
\bibinfo{author}{\bibfnamefont{M.~A.} \bibnamefont{Nielsen}} \bibnamefont{and}
  \bibinfo{author}{\bibfnamefont{I.~L.} \bibnamefont{Chuang}},
  \emph{\bibinfo{title}{Quantum computation and quantum information}}
  (\bibinfo{publisher}{Cambridge University Press}, \bibinfo{year}{2000}).

\bibitem[{\citenamefont{Devoret et~al.}(2004)\citenamefont{Devoret, Wallraff,
  and Martinis}}]{Devoret04}
\bibinfo{author}{\bibfnamefont{M.~H.} \bibnamefont{Devoret}},
  \bibinfo{author}{\bibfnamefont{A.}~\bibnamefont{Wallraff}}, \bibnamefont{and}
  \bibinfo{author}{\bibfnamefont{J.~M.} \bibnamefont{Martinis}},
  \bibinfo{journal}{cond-mat/0411174}  (\bibinfo{year}{2004}).

\bibitem[{\citenamefont{Wendin and Shumeiko}(2005)}]{Wendin05}
\bibinfo{author}{\bibfnamefont{G.}~\bibnamefont{Wendin}} \bibnamefont{and}
  \bibinfo{author}{\bibfnamefont{V.}~\bibnamefont{Shumeiko}},
  \bibinfo{journal}{cond-mat/0508729}  (\bibinfo{year}{2005}).

\bibitem[{\citenamefont{Berkley et~al.}(2003)\citenamefont{Berkley, Xu, Ramos,
  Gubrud, Strauch, Johnson, Anderson, Dragt, Lobb, and Wellstood}}]{Berkley03}
\bibinfo{author}{\bibfnamefont{A.~J.} \bibnamefont{Berkley \textit{et al.}}},
  \bibinfo{journal}{Science} \textbf{\bibinfo{volume}{300}},
  \bibinfo{pages}{1548} (\bibinfo{year}{2003}).

\bibitem[{\citenamefont{Pashkin et~al.}(2003)\citenamefont{Pashkin, Yamamoto,
  Astafiev, Nakamura, Averin, and Tsai}}]{Pashkin03}
\bibinfo{author}{\bibfnamefont{Y.~A.} \bibnamefont{Pashkin \textit{et al.}}},
  \bibinfo{journal}{Nature} \textbf{\bibinfo{volume}{421}},
  \bibinfo{pages}{823} (\bibinfo{year}{2003}).

\bibitem[{\citenamefont{Majer et~al.}(2005)\citenamefont{Majer, Paauw, ter
  Haar, Harmans, and Mooij}}]{Majer04}
\bibinfo{author}{\bibfnamefont{J.~B.} \bibnamefont{Majer \textit{et al.}}},
  \bibinfo{journal}{Phys. Rev. Lett.} \textbf{\bibinfo{volume}{94}},
  \bibinfo{pages}{090501} (\bibinfo{year}{2005}).

\bibitem[{\citenamefont{{Chiorescu} et~al.}(2004)\citenamefont{{Chiorescu},
  {Bertet}, {Semba}, {Nakamura}, {Harmans}, and {Mooij}}}]{Chiorescu04}
\bibinfo{author}{\bibfnamefont{I.}~\bibnamefont{{Chiorescu \textit{et al.}}}},
 \bibinfo{journal}{Nature (London)}
  \textbf{\bibinfo{volume}{431}}, \bibinfo{pages}{159} (\bibinfo{year}{2004}).

\bibitem[{\citenamefont{McDermott et~al.}(2005)\citenamefont{McDermott,
  Simmonds, Steffen, Cooper, Cicak, Osborn, Oh, Pappas, and
  Martinis}}]{McDermott04}
\bibinfo{author}{\bibfnamefont{R.}~\bibnamefont{McDermott \textit{et al.}}},
 \bibinfo{journal}{Science}
  \textbf{\bibinfo{volume}{307}}, \bibinfo{pages}{1299} (\bibinfo{year}{2005}).

\bibitem[{\citenamefont{Yamamoto et~al.}(2003)\citenamefont{Yamamoto, Pashkin,
  Astafiev, Nakamura, and Tsai}}]{Yamamoto03}
\bibinfo{author}{\bibfnamefont{T.}~\bibnamefont{Yamamoto \textit{et al.}}},
  \bibinfo{journal}{Nature} \textbf{\bibinfo{volume}{425}},
  \bibinfo{pages}{941} (\bibinfo{year}{2003}).

\bibitem[{\citenamefont{Steffen et~al.}(2006)\citenamefont{Steffen, Ansmann,
  Bialczak, Katz, Lucero, McDermott, Neeley, Weig, Cleland, and
  Martinis}}]{Steffen06}
\bibinfo{author}{\bibfnamefont{M.}~\bibnamefont{Steffen \textit{et al.}}},
 \bibinfo{journal}{Science}
  \textbf{\bibinfo{volume}{313}}, \bibinfo{pages}{1423} (\bibinfo{year}{2006}).

\bibitem[{\citenamefont{Makhlin et~al.}(1999)\citenamefont{Makhlin, Sch{\"o}n,
  and Shnirman}}]{Makhlin99}
\bibinfo{author}{\bibfnamefont{Y.}~\bibnamefont{Makhlin}},
  \bibinfo{author}{\bibfnamefont{G.}~\bibnamefont{Sch{\"o}n}},
  \bibnamefont{and} \bibinfo{author}{\bibfnamefont{A.}~\bibnamefont{Shnirman}},
  \bibinfo{journal}{Nature} \textbf{\bibinfo{volume}{398}},
  \bibinfo{pages}{305} (\bibinfo{year}{1999}).

\bibitem[{\citenamefont{Blais et~al.}(2004)\citenamefont{Blais, Huang,
  Wallraff, Girvin, and Schoelkopf}}]{Blais04}
\bibinfo{author}{\bibfnamefont{A.}~\bibnamefont{Blais \textit{et al.}}},
 \bibinfo{journal}{Phys. Rev. A}
  \textbf{\bibinfo{volume}{69}}, \bibinfo{pages}{062320}
  (\bibinfo{year}{2004}).

\bibitem[{\citenamefont{Blais et~al.}(2006)\citenamefont{Blais, Gambetta,
  Wallraff, Schuster, Girvin, Devoret, and Schoelkopf}}]{Blais06}
\bibinfo{author}{\bibfnamefont{A.}~\bibnamefont{Blais \textit{et al.}}},
  \bibinfo{journal}{condmat/0612038}
  (\bibinfo{year}{2006}).

\bibitem[{\citenamefont{Wallraff et~al.}(2004)\citenamefont{Wallraff, Schuster,
  Blais, Frunzio, Huang, Majer, Kumar, Girvin, and Schoelkopf}}]{wallraff04c}
\bibinfo{author}{\bibfnamefont{A.}~\bibnamefont{Wallraff \textit{et al.}}},
 \bibinfo{journal}{Nature (London)}
  \textbf{\bibinfo{volume}{431}}, \bibinfo{pages}{162} (\bibinfo{year}{2004}).

\bibitem[{\citenamefont{xi~Liu et~al.}(2005{\natexlab{a}})\citenamefont{xi~Liu,
  Wei, Tsai, and Nori}}]{liu05a}
\bibinfo{author}{\bibfnamefont{Y.}~\bibnamefont{xi~Liu \textit{et al.}}},
  (\bibinfo{year}{2005}{\natexlab{a}}), \bibinfo{note}{cond-mat/0509236}.

\bibitem[{\citenamefont{Johansson et~al.}(2006)\citenamefont{Johansson, Saito,
  Meno, Nakano, Ueda, Semba, and Takayanagi}}]{Johansson06}
\bibinfo{author}{\bibfnamefont{J.}~\bibnamefont{Johansson \textit{et al.}}},
  \bibinfo{journal}{Phys. Rev. Lett.} \textbf{\bibinfo{volume}{96}},
  \bibinfo{pages}{127006} (\bibinfo{year}{2006}).

\bibitem[{\citenamefont{Andre et~al.}(2006)\citenamefont{Andre, DeMille, Doyle,
  Lukin, Maxwell, Rabl, Schoelkopf, and Zoller}}]{Andre06}
\bibinfo{author}{\bibfnamefont{A.}~\bibnamefont{Andre \textit{et al.}}},
  \bibinfo{journal}{Nature Physics} \textbf{\bibinfo{volume}{2}},
  \bibinfo{pages}{636} (\bibinfo{year}{2006}).

\bibitem[{\citenamefont{Schuster et~al.}(2005)\citenamefont{Schuster, Wallraff,
  Blais, Frunzio, Huang, Majer, Girvin, and Schoelkopf}}]{Schuster05}
\bibinfo{author}{\bibfnamefont{D.~I.} \bibnamefont{Schuster \textit{et al.}}},
 \bibinfo{journal}{Phys. Rev. Lett.}
  \textbf{\bibinfo{volume}{94}}, \bibinfo{pages}{123602}
  (\bibinfo{year}{2005}).

\bibitem[{\citenamefont{Wallraff et~al.}(2005)\citenamefont{Wallraff, Schuster,
  Blais, Frunzio, Majer, Devoret, Girvin, and Schoelkopf}}]{wallraff05}
\bibinfo{author}{\bibfnamefont{A.}~\bibnamefont{Wallraff \textit{et al.}}},
  \bibinfo{journal}{Phys. Rev. Lett.}
  \textbf{\bibinfo{volume}{95}}, \bibinfo{eid}{060501} (\bibinfo{year}{2005}).

\bibitem[{\citenamefont{Bouchiat et~al.}(1998)\citenamefont{Bouchiat, Vion,
  Joyez, Esteve, and Devoret}}]{Bouchiat98}
\bibinfo{author}{\bibfnamefont{V.}~\bibnamefont{Bouchiat \textit{et al.}}},
  \bibinfo{journal}{Physica Scripta} \textbf{\bibinfo{volume}{T76}},
  \bibinfo{pages}{165} (\bibinfo{year}{1998}).

\bibitem[{\citenamefont{Vion et~al.}(2002)\citenamefont{Vion, Aassime, Cottet,
  Joyez, Pothier, Urbina, Esteve, and Devoret}}]{Vion02}
\bibinfo{author}{\bibfnamefont{D.}~\bibnamefont{Vion \textit{et al.}}},
  \bibinfo{journal}{Science} \textbf{\bibinfo{volume}{296}},
  \bibinfo{pages}{886} (\bibinfo{year}{2002}).

\bibitem[{\citenamefont{Frunzio et~al.}(2005)\citenamefont{Frunzio, Wallraff,
  Schuster, Majer, and Schoelkopf}}]{Frunzio05}
\bibinfo{author}{\bibfnamefont{L.}~\bibnamefont{Frunzio \textit{et al.}}},
  \bibinfo{journal}{IEEE Trans. Appl. Supercond.}
  \textbf{\bibinfo{volume}{15}}, \bibinfo{pages}{860} (\bibinfo{year}{2005}).

\bibitem[{\citenamefont{xi~Liu et~al.}(2005{\natexlab{b}})\citenamefont{xi~Liu,
  You, Wei, Sun, and Nori}}]{liu05b}
\bibinfo{author}{\bibfnamefont{Y.}~\bibnamefont{xi~Liu \textit{et al.}}},
  \bibinfo{journal}{Physical Review Letters} \textbf{\bibinfo{volume}{95}},
  \bibinfo{pages}{087001} (\bibinfo{year}{2005}{\natexlab{b}}).

\bibitem[{\citenamefont{Ithier et~al.}(2005)\citenamefont{Ithier, Collin,
  Joyez, Meeson, Vion, Esteve, Chiarello, Shnirman, Makhlin, Schriefl
  et~al.}}]{Ithier05}
\bibinfo{author}{\bibfnamefont{G.}~\bibnamefont{Ithier \textit{et al.}}},
  \bibnamefont{et~al.}, \bibinfo{journal}{Phys. Rev. B}
  \textbf{\bibinfo{volume}{72}}, \bibinfo{pages}{134519}
  (\bibinfo{year}{2005}).

\bibitem[{\citenamefont{Raimond et~al.}(2001)\citenamefont{Raimond, Brune, and
  Haroche}}]{raimond01}
\bibinfo{author}{\bibfnamefont{J.}~\bibnamefont{Raimond}},
  \bibinfo{author}{\bibfnamefont{M.}~\bibnamefont{Brune}}, \bibnamefont{and}
  \bibinfo{author}{\bibfnamefont{S.}~\bibnamefont{Haroche}},
  \bibinfo{journal}{Rev. Mod. Phys.} \textbf{\bibinfo{volume}{73}},
  \bibinfo{pages}{565} (\bibinfo{year}{2001}).

\bibitem[{\citenamefont{Schmidt-Kaler et~al.}(2003)\citenamefont{Schmidt-Kaler,
  Haffner, Riebe, Gulde, Lancaster, Deuschle, Becher, Roos, Eschner, and
  Blatt}}]{schmidt-kaler03}
\bibinfo{author}{\bibfnamefont{F.}~\bibnamefont{Schmidt-Kaler \textit{et al.}}},
  \bibinfo{journal}{Nature} \textbf{\bibinfo{volume}{422}},
  \bibinfo{pages}{408} (\bibinfo{year}{2003}).

\end{thebibliography}
\end{document}